\documentclass{osa-article}

\journal{osac}

\newcommand{\comment}[1]{\ignorespaces}

\articletype{Research Article}

\usepackage{graphicx}
\usepackage{dcolumn}
\usepackage{bm}
\usepackage{ulem}
\usepackage{lineno}

\begin{document}
	
	\title{Cloaked near-field probe for non-invasive near-field optical microscopy}
	
	\author{Felipe Bernal Arango,\authormark{1} Filippo Alpeggiani,\authormark{1} Donato Conteduca,\authormark{2} Aron Opheij,\authormark{1} Aobo Chen,\authormark{4} Mohamed I. Abdelrahman,\authormark{4} Thomas Krauss,\authormark{2} Andrea Al\`u,\authormark{3} Francesco Monticone,\authormark{4,$\dagger$} L. Kuipers\authormark{1,*}}
	
	\address{\authormark{1}Department of Quantum Nanoscience, Delft University of Technology, Delft, The Netherlands\\
	\authormark{2}Department of Physics, University of York, York, United Kingdom\\
	\authormark{3}Advanced Science Research Center,	City University of New York, New York, NY, 10031, USA\\
	\authormark{4}School of Electrical and Computer Engineering, Cornell University, Ithaca, NY, 14853, USA}

	\email{\authormark{$\dagger$}francesco.monticone@cornell.edu} 
	\email{\authormark{*}l.kuipers@tudelft.nl} 

	\begin{abstract}
		Near-field scanning optical microscopy is a powerful technique for imaging below the diffraction limit, which has been extensively used in bio-medical imaging and nanophotonics. However, when the electromagnetic fields under measurement are strongly confined, they can be heavily perturbed by the presence of the near-field probe itself. Here, taking inspiration from scattering-cancellation invisibility cloaks, Huygens-Kerker scatterers, and cloaked sensors, we design and fabricate a cloaked near-field probe. We show that, by suitably nanostructuring the probe, its electric and magnetic polarizabilities can be controlled and balanced. As a result, probe-induced perturbations can be largely suppressed, effectively cloaking the near-field probe without preventing its ability to measure. We experimentally demonstrate the cloaking effect by comparing the interaction of conventional and nanostructured probes with a representative nanophotonic structure, namely, a 1D photonic-crystal cavity. Our results show that, by engineering the structure of the probe, one can systematically control its back-action on the resonant fields of the sample and decrease the perturbation by $>$70\% with most of our modified probes, and by up to one order of magnitude for the best probe, at probe-sample distances of 100 nm. Our work paves the way for non-invasive near-field optical microscopy of classical and quantum nano-systems.
	\end{abstract}

	\section{Introduction}
	Measurement implies perturbation. This is a well established fact both in quantum and classical physics, and it becomes particularly evident at scales comparable to the wavelength of the physical elements (particles, fields) under measurement. Indeed, for electromagnetic fields the so-called optical theorem \cite{jackson1999,bohren2008absorption} implies that, if the total scattering cross section of a passive sensor is identically zero, then no energy can be absorbed, i.e., measured, since the extinction and absorption cross section of the sensor are also required to be zero. Hence, any sensor is required to scatter, thus perturbing the fields under study, in order to measure them. Rather counter-intuitively, in this paper we experimentally demonstrate a sensing element, an optical near-field probe, that is able to measure an electromagnetic field distribution with minimized perturbations.\par
	
	The idea of using a small probe in the near-field of an illuminated sample to achieve optical imaging not limited by diffraction dates back to the work of Synge in the 1920s~\cite{Synge_1928}. In recent decades, thanks to tremendous advances in nanotechnology, near-field scanning optical microscopy (NSOM) is becoming an essential tool for probing subdiffractive features of nanostructures in biological systems (e.g., proteins on the surface of a cell~\cite{deLange2001}) and nanophotonic structures~\cite{Dorfmuller_2011, Schnell2009}. Clearly, this technique induces perturbations on the field distributions under measurement, due to the scattering from the probe itself, consistent with our discussion above. While for structures with a large photonic-mode volume this perturbation is negligible \cite{Balistreri_01}, for highly-confined resonant field distributions -- as common for nanophotonic structures -- it becomes particularly severe, leading to significant image artifacts, because the effective volume of the supported photonic modes may be comparable to the effective size of the near-field probe, given by its scattering cross section (and proportional to its polarizability). Indeed, it was predicted by Koenderink et al.~\cite{Koenderink_2005} that the interaction between an NSOM probe and the resonant near-field distribution of a photonic cavity could spectrally shift the photonic modes of the system. While this effect provides a useful mechanism to finely tune the cavity resonance and can be used to determine the complex mode volumes of quasi-normal cavity modes~\cite{Cognee_2019}, it represents an obstacle for imaging resonant nanophotonic structures, as the field measured by the probe can be heavily distorted by the probe itself. These perturbation effects were subsequently observed experimentally in a large body of work \cite{Lalouat_2007,Intoti_2008,Vignolini_2010,Burresi_2010}.\par 
	
	Suppressing the perturbing effect of an NSOM probe is an appealing goal that paves the way for minimally perturbative subdiffractive optical imaging and sensing of highly resonant structures. Low-invasive probes may enable near-field measurements on fragile states that would otherwise be disrupted by the probe scattering, which is of particular relevance in cavity quantum electrodynamics or in the study of single emitters and their interaction with nanostructures without the issue of probe-induced Lamb-shifts. Over the past few decades, several strategies have been proposed to make optical scatterers invisible by minimizing their scattering cross section \cite{Fleury_2015} using suitably designed plasmonic cloaks \cite{Kerker_1975, Alu_2008}, metasurfaces \cite{Estakhri_2014,Ni_2015}, or complex metamaterial shells \cite{Schurig_06,Alu_2005,Valentine_2009,Soric_14,chen2019active}. Most importantly for our purposes, it was theoretically demonstrated in \cite{Alu_2009} that ``cloaked sensors'' could be designed with minimized scattering cross sections without preventing their ability to absorb (i.e., measure) a finite amount of energy. In fact, while the optical theorem imposes constraints on the scattering and absorption properties of any passive scattering system, it does not set a lower limit on the \textit{ratio} between scattering and absorption cross section, which implies that almost-invisible objects may absorb a small, yet non-zero, amount of energy \cite{Soric_14,Fleury_14}. The potential of these ideas to realize invisible near-field probes was theoretically investigated in \cite{Alu_2010,Chen_2017}. In particular, it was recognized that the cloaking mechanism should not ``shield'' the probe from the external fields, as it is crucial that a sufficient amount of light is still coupled into the probe, in order to guarantee a detectable level of signal. Scattering-cancellation techniques \cite{Alu_2005,Alu_2008,Alu_2009,Alu_2010}, based on compensating the induced dipole moments in a scatterer with counter-oscillating dipole moments in the cloak, are ideally suited for this purpose, as they allow reducing the scattered field without shielding the object to be concealed inside an impenetrable shell. In this way, not only the internal fields are not reduced, they can actually be enhanced \cite{Alu_2009,Monticone_2014}. Moreover, for many applications, it may not even be necessary to minimize the total scattering cross section of the sensor. Indeed, if the sensor's position relative to the sample under measurement is known, it would be more than sufficient to suppress the scattered fields in that specific direction (typically the backward direction). This can be achieved through interference effects between electric and magnetic dipole scattering (or higher-order scattering contributions), following for example the well-established Kerker's conditions for zero back-scattering \cite{Kerker_1975,Alukerker_2010,picardi2018janus,picardi2019experimental}, which represents a particularly appealing solution for our purposes, as further discussed in the next section. \par
	Inspired by these ideas, in this work we design and fabricate a ``cloaked'' NSOM probe, and we experimentally demonstrate its sensing performance by studying the probe interaction with a photonic-crystal cavity \cite{Conteduca_2017}. 
	
	\section{Results}
\begin{figure}[htb!]
		\centering{\includegraphics[width=0.6\textwidth]{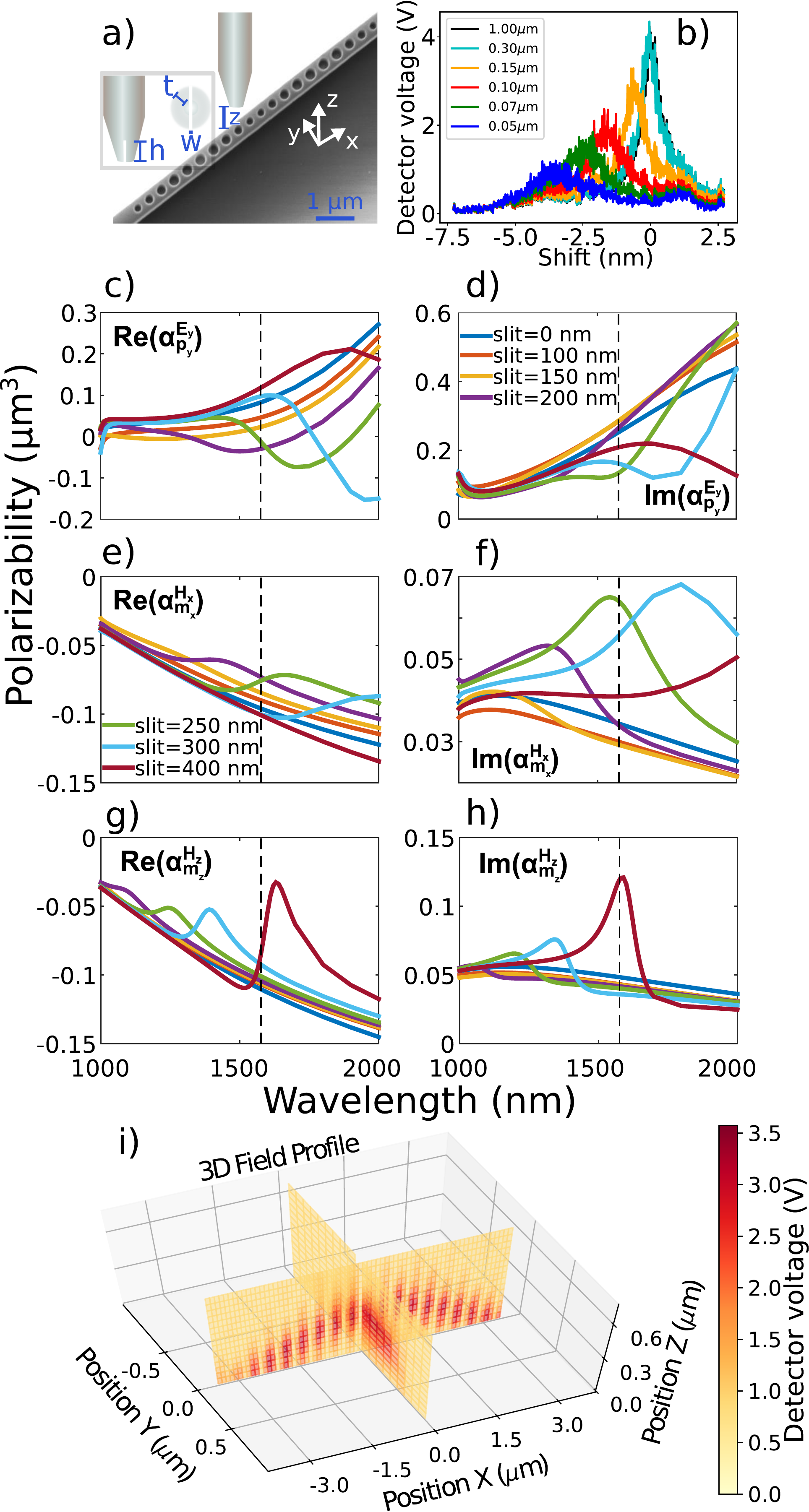}}
		\caption{Description of the measurements and the perturbation problem in NSOM for confined fields. (a) SEM picture of a 1D photonic crystal cavity together with a sketch of an aperture-type NSOM probe. The inset includes a sketch of a nanostructured probe with slits carved along the aluminum shell. The slit is aligned along the x axis. (b) Example of measured transmission spectra through the cavity, clearly showing the change in amplitude and width due to the presence of a conventional, unmodified probe at different heights. (c-h) Retrieved real and imaginary parts of the polarizability components~\cite{Bernal_2013,Bernal_2014} of the near-field probe, $\alpha^{E_{y}}_{p_{y}}$, $\alpha^{H_{x}}_{m_{x}}$ and $\alpha^{H_{z}}_{m_{z}}$, as a function of the wavelength and slit length. The dashed line indicates the position of the cavity resonance at 1576.2~nm. (i) Near-field amplitude map created by measuring the transmission through a conventional near-field aperture probe while 3D-raster-scanning the probe. The figure shows a horizontal X-Y plane measured in shear force mode, a vertical Y-Z plane measured at X$=0$, and a Z-X plane measured at Y$=0$.
		}\label{mappingFields}
	\end{figure}

	In aperture-based near-field probes, the tip of a tapered optical fiber is coated with aluminum, and subsequently an aperture is opened at the tip. The subwavelength aperture interacts with the evanescent near-fields of the sample, coupling them into the guided modes of the tapered optical fiber, in which the fields are guided towards a detector as the probe is scanned above an illuminated sample (See Fig.~\ref{mappingFields}(a)). While great progress has been made in the design and operation of apertureless NSOM probes, including dielectric probes, aperture-based NSOM probes offer the important advantage of illuminating the sample and/or collecting light very locally, thus creating images that are essentially background-free \cite{Novotny2006,fleischer2012near}. This comes at the cost of a larger probe size, which usually implies larger scattering, and therefore stronger scattering-induced artifacts. Our goal in this work is to overcome this trade-off. 
	
	Although aperture tips are larger than apertureless ones, their size is still typically much smaller than the wavelength. Thus, the interaction with the sample field, and the consequent scattering process, are dominated by the induced electric dipole moment $\bf{p}$ and magnetic dipole moment $\bf{m}$ on the probe, i.e. $\left(\bf{p};\bf{m}\right)=\bar{\bf{\alpha}} \left(\bf{E};\bf{H}\right)$, where $\bar{\bf{\alpha}}$ is the magneto-electric polarizability tensor \cite{Sersic_2009,Bernal_2013} of the probe: $\bar{\bf{\alpha}}=\left(\bar{\bf{\alpha}}_{E},\bar{\bf{\alpha}}_{EH};\bar{\bf{\alpha}}_{HE},\bar{\bf{\alpha}}_{H}\right)$. The upper diagonal block $\bar {\alpha }_{E}$ is the electric polarizability tensor, the lower diagonal block $\bar {\alpha }_H$ is the magnetic polarizability tensor, and the off-diagonal blocks $\bar{\bf{\alpha}}_{EH}$ and $\bar{\bf{\alpha}}_{HE}$ hold the magneto-electric response of the probe, i.e., the electric (magnetic) moment induced by magnetic (electric) field; $\left[\bf{E};\bf{H}\right]$ are the electric and magnetic fields at the probe location. As shown in~\cite{Olmon_2010,Devaux_2000, Kihm_2011}, it is indispensable for an accurate description of the NSOM probe response to take into account the magnetic field interactions. The source of the scattered field is the induced current on the probe, which is proportional to the dipole moments~\cite{Ribaric_95}. If magneto-electric coupling can be neglected, the electric and magnetic polarizability tensors, $\bar {\alpha }_{E}$ and $\bar {\alpha }_{H}$, can be interpreted as admittance and impedance tensors, respectively, relating the local fields to the induced currents in an equivalent circuit model of the scatterer. Thus, an object with an element of the electric polarizability tensor, e.g., $\alpha^{E_{y}}_{p_{y}}$, having positive (negative) real part is seen as a local shunt capacitance (inductance) by the local electric field $E_{y}$. Similarly, an element of the magnetic polarizability tensor with positive (negative) real part corresponds to a local series capacitance (inductance). In particular, aperture probes of this type usually have stronger inductive impedance due to their magnetic response~\cite{Burresi_2010}. The stronger inductive impedance is evidenced in Fig. 1(b), where we show spectra of transmission measurements of a photonic crystal cavity as a function of the probe-cavity distance \textit{z} (See Fig. 1(a)) for a conventional unmodified probe. In these spectra we observe how the resonance is blue shifted as the probe approaches the cavity, and the width of the resonance is increased due to loss channels caused by scattering.\par
	Our design of a cloaked NSOM probe is heuristically based on this interpretation in terms of impedance/admittance and on a general strategy to control them. Indeed, as proposed in \cite{Chen_2017}, if slits are carved along the probe (parallel to the probe axis, as shown in the inset of Fig.~\ref{mappingFields}(a)), such slits would act as short transmission-line segments whose length can be used to control the overall impedance of the probe and possibly induce resonances in the polarizability tensor elements. We also note that these resonances can be further tuned by varying the thickness of the aluminum coating, which also modifies the relative impedance/admittance of the probe. In addition, a change of material properties, for instance the presence of a layer of Al$_{3}$O${_2}$, whether by natural or deliberate oxidation, is another option to tune the resonances. We also note that various forms of nano-structuring of an NSOM probe have been investigated, computationally and experimentally, in the recent literature (e.g., \cite{fleischer2012near,smajic2011numerical,qian2015plasmonic}, and references therein), but for different purposes unrelated to scattering suppression. \par
	In Fig.~\ref{mappingFields}(c-h), we plot all the numerically calculated electric and magnetic polarizability tensor elements for different slit lengths, demonstrating the presence of a slit-induced resonance in both the electric and magnetic response of this nanostructured probe. Such slit-induced resonance allows us to tune the different elements of the magneto-electric polarizability tensor to a large degree, so that we can either increase or decrease the real/imaginary part of the magnetic/electric polarizability.\par 
	In order to quantify the impact of our nanostructured probe on the near-field measurements of a nanophotonic structure, we turn to perturbation theory, which  tells us that the wavelength shift induced by the probe on the cavity modes is related to the change in energy induced by the perturbing element~\cite{Koenderink_2005}. When we neglect the magneto-electric cross-coupling terms $\bar{\bf{\alpha}}_{EH}$ and $\bar{\bf{\alpha}}_{HE}$, which are typically small, and the linewidth change due to the loss channels introduced by the perturbation, the local resonance shift induced by the probe is approximately given by \cite{Koenderink_2005, Burresi_2010}:
	
	\begin{equation}\label{equationshift}
		\frac{\Delta \lambda_{r}(\bm{r})}{\lambda_{r0}}\approx (Re(\bar{\alpha}_{\mbox{\tiny{E}}}) \bm{E}(\bm{r}))\cdot \frac{\bm{E}^{*}(\bm{r})}{U_0} +(Re({\bar{\alpha}}_{\mbox{\tiny{H}}}) \bm{H}(\bm{r}))\cdot \frac{ \bm{H} ^{*}(\bm{r})}{U_0}.
	\end{equation}
	
	Here, $\lambda_{r0}$ is the wavelength of the unperturbed resonant mode, $\Delta \lambda_{r}(\bf{r})$ is the local change in resonant wavelength due to the perturbation, $U_{0}$ is the energy stored in the electromagnetic fields of the unperturbed cavity and $\bm{r}$ is the position of the probe. As seen in Fig. 1(c,e,g), when driving the probe far from resonance, the real electric polarizability term is positive (capacitive), which therefore increases the energy of the system and red-shifts the cavity resonance to longer wavelengths, while the real magnetic polarizability term is negative (inductive), hence blue-shifting the cavity resonance. Whether the blue- or red-shift dominates depends on the specific probe geometry and material composition.\par
	
	Based on these considerations, it is therefore clear from Eq.~(\ref{equationshift}) that we have two options to minimize the probe-induced perturbation: we can either minimize the dominant terms of both $\bar {\alpha }_{E}$ and $\bar {\alpha }_{H}$ simultaneously, or make them equal and opposite, thus compensating their wavelength shift. However, by inspecting Fig.~\ref{mappingFields}(c,e,g), it is evident that the first option is not available, as the slit-induced resonance allows $\bar {\alpha }_{E}$ to become zero, but $\bar {\alpha }_{H}$ cannot. Instead, the second option, with $Re(\bar{\alpha }_{E})=-Re(\bar {\alpha }_{H})$, can be easily achieved, producing a net zero wavelength shift. Moreover, from Fig.~\ref{mappingFields}(f), it is also clear that the slits induce an increase in the imaginary part of $\alpha^{H_{x}}_{m_{x}}$, making it comparable to the other real and imaginary polarizability values. In this way, we are able to approach an interesting condition for the in-plane polarizabilities, $\alpha^{E_{y}}_{p_{y}} \approx -i \, \alpha^{H_{x}}_{m_{x}}$ (approached, e.g., by the cyan curve, in Figs.~\ref{mappingFields}(c)-(f), for 300 nm slits, near the cavity resonance), which implies that the induced in-plane electric and magnetic dipoles are in-phase when this condition is respected, which in turn implies that the back-scattering into the sample is minimized, similar to the Huygens-Kerker condition~\cite{Kerker_1975,Alukerker_2010,picardi2018janus,picardi2019experimental}, but for near-field driving of the scatterer with evanescent $\bm{E}$ and $\bm{H}$ fields dephased by 90$^{\circ}$ \cite{wei2018directional} (the standard Huygens-Kerker condition would instead require $\alpha^{E_{y}}_{p_{y}} \approx \alpha^{H_{x}}_{m_{x}}$ for in-phase driving fields as those of a propagating plane wave). 
	Thus, our goal is to realize a nanostructured NSOM probe with balanced electric and magnetic responses, such that the back-scattering into the sample, as well as the total energy change and therefore the resonance shift, are minimized, effectively cloaking the probe. In this way, as demonstrated in the following, we are able to access an operation mode in which we measure a very close approximation of the actual eigenmode of the photonic system under study and not the mode of the perturbed system.\par
	
	To test our ability to cloak a near-field probe based on this strategy, we use as a benchmark a 1D photonic-crystal cavity (Fig.~\ref{mappingFields}(a)), fabricated in SOI with a TE (transverse electric) resonant mode centered around 1550~nm, a Q value of $\sim$3000, and a mode volume of $\sim1\times10^{-19}$~$m^3$. We image the field profile of this nanophotonic cavity using a homebuilt, polarization-sensitive NSOM~\cite{Feber_2013}, mapping the amplitude of the fields in 3D when driving the cavity near resonance (at a wavelength of 1576.2~nm). The 3D plot in Fig.~\ref{mappingFields}(i) shows three planes of this field map. In the horizontal XY plane as well as in the vertical XZ plane we see a periodic repetition of, respectively, horizontal and vertical red features, corresponding to the field antinodes or maxima, which is a clear signature of the standing-wave pattern of the cavity mode. We also observe how the amplitude of the fields decreases as the distance from the surface increases, as expected.\par
	
	Importantly, we see in Fig.~\ref{mappingFields}(i) that, close to the center of the cavity and above the surface (i.e., for $X,Y,Z=0$), the measured field amplitude is significantly lower, which is surprising since, for this type of cavities, the highest field amplitude is expected to be at the center of the cavity \cite{Conteduca_2017}. This zone of unexpectedly low field amplitudes indicates that the probe is exerting a strong perturbation on the measurements. Indeed, consistent with Eq.~(\ref{equationshift}), a strong perturbation induces a large wavelength shift in the resonant mode, especially in spatial regions with high field intensity. This determines a subsequent decrease of intensity in the perturbed cavity, due to the cavity being out of resonance with respect to the wavelength of the driving laser, which in turn results in a reduced detected intensity. We quantify the perturbation introduced by the probe by measuring the transmission spectra of the 1D photonic-crystal cavity as a function of the position of the probe, which allows us to determine the wavelength shift in Eq.~(\ref{equationshift}) (See Appendix). These transmission measurements are performed simultaneously with the collection through the aperture probe, so that we can easily compare the impact of the probe scattering on both the near-field probe-based measurements and the transmission measurements.\par
	
	\begin{figure}[htb!]
		\centering{\includegraphics[width=0.75\textwidth]{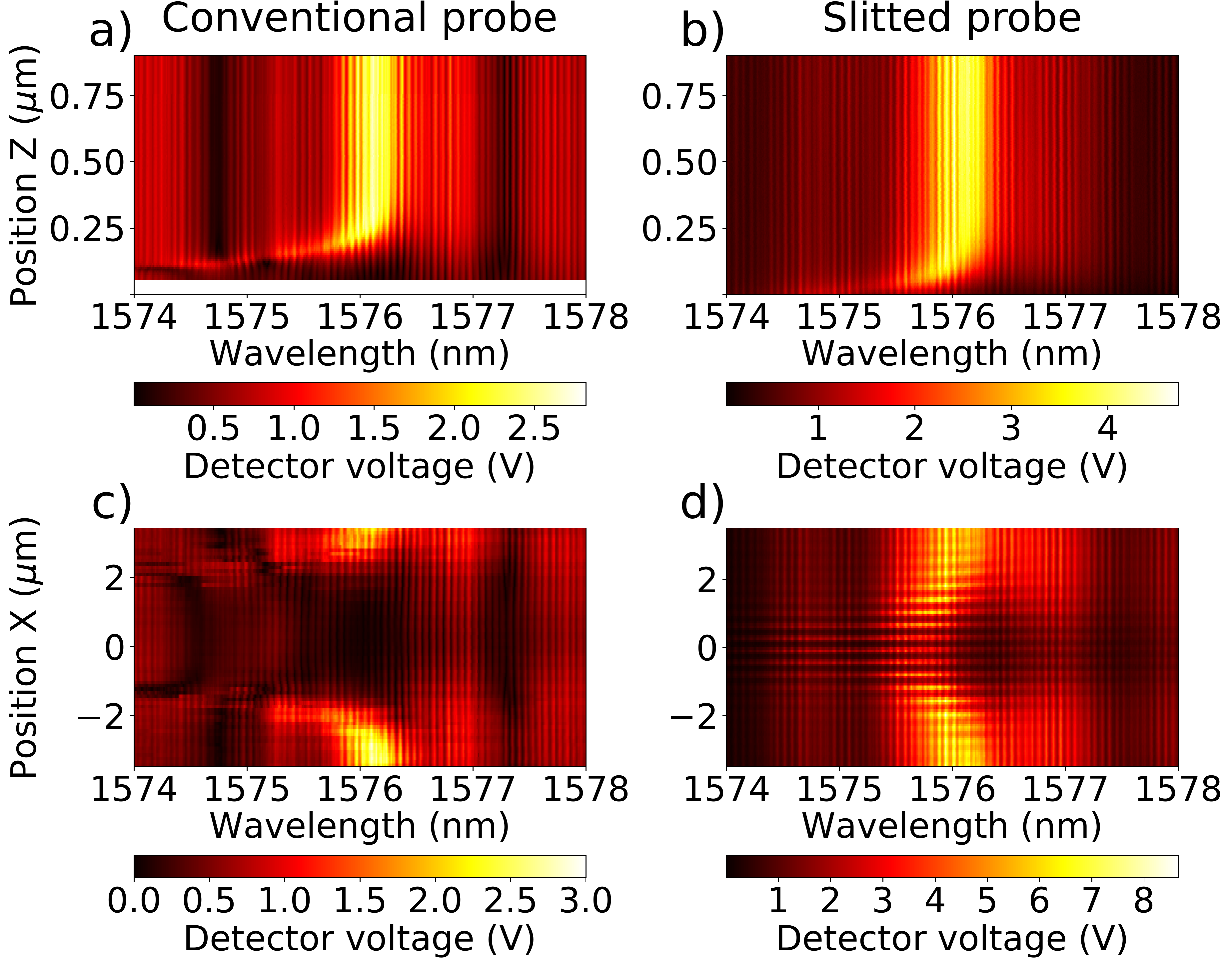}}
		\caption{The top images (panels (a) and (b)) present transmission spectral measurements for the nanophotonic cavity in Fig.~\ref{mappingFields}, with (a) a conventional or (b) a cloaked probe (Probe 3 in Appendix) approaching the center of the cavity from the top. The two bottom images ((c) and (d)) present the transmission maps with probes in contact with the sample and approaching the center of the cavity along its mirror-symmetry axis.}\label{figure2shifts}
	\end{figure}
	
	In Fig.~\ref{figure2shifts}, we present measurements for both a conventional probe and a slitted probe designed to have balanced electric and magnetic polarizabilities, as discussed above. We perform transmission spectroscopy measurements following two different probe trajectories towards the center of the cavity. In the first trajectory (Fig.~\ref{figure2shifts}(a)~and~(b)), we move the probe vertically ($z$-direction) from the top-center of the cavity and, while measuring the transmission spectra, we also measure the change in probe-sample distance with a SmarAct GmbH PicoScale interferometer. In the second trajectory, we measure transmission as we move the probe towards the center of the cavity along its axis of mirror symmetry ($x$-axis), while keeping the probe at close proximity ($\sim$10~nm, see Appendix) to the sample (Fig.~\ref{figure2shifts}(c)~and~(d)).\par
	Fig.~\ref{figure2shifts}(a) shows that, when a conventional NSOM probe is far from the cavity, there is a clear resonance in the transmission spectrum centered around 1576.2~nm. As the probe gets closer to the center of the cavity, however, this resonance starts shifting towards shorter wavelengths. The magnitude of the resonance shift is a quantitative measure of the perturbation introduced by the probe. At a probe height of around 110~nm, the resonance is centered around 1574~nm, namely, the probe induces a resonance blue-shift of 2.2~nm, which is qualitatively consistent with our numerical calculations in Fig. 1(c,e,g), where we see that a probe with no slits has a dominant magnetic polarizability (its real part is 16\% stronger than the real part of the electric polarizability, at 1576.2~nm).
	A similar behaviour is found for the trajectory along the cavity axis. As seen in Fig.~\ref{figure2shifts}(c), the cavity resonance is clearly visible when a conventional NSOM probe is far ($\pm~3~\mu$m) from the cavity center, where the cavity fields are less intense. As the probe approaches the center ($X=0$), we clearly see how the resonance is strongly blue shifted, broadened, and eventually completely disrupted.\par
	In stark contrast, for a slitted probe with balanced electric-magnetic response, our measurements in Fig.~\ref{figure2shifts}(b) show a significantly smaller blue-shift of the cavity resonance.
	The wavelength shift is decreased by an order of magnitude at a probe height of 110~nm, from 2.2~nm to about 0.2~nm. Furthermore, this small shift is less than 2/3 of the resonance FWHM (0.38~nm) at around 50~nm height, which is a practical distance for near-field measurements.
	The reduced perturbation introduced by the slitted probe is even more evident considering the horizontal trajectory along the cavity axis, as shown in Fig.~\ref{figure2shifts}(d). When the probe is far from the center of the cavity, we again see the resonance centered at 1576.2~nm, as in Fig.~\ref{figure2shifts}(c); however, when the slitted probe is moved to the center of the cavity, we now observe only a very small shift of 0.5~nm, and reduced resonance broadening. The fact that the perturbation is reduced to such a degree, even for a probe at a distance of only 10~nm from the field hot-spot at the center of the cavity, clearly demonstrates the ``cloaking'' effect in action.\par 
	Another feature visible in all the measurements in Fig.~\ref{figure2shifts} are the thin (FWHM~0.02~nm) vertical dark-bright lines separated by 0.06~nm, which are the signature of Fabry-Perot resonances created between the waveguide facets of the chip and the photonic-crystal cavity. Interestingly, we note that these Fabry-Perot features are not perturbed by the presence of either conventional or nanostructured probes. This is expected, since the large mode volume of these Fabry-Perot modes makes them insensitive to the comparatively small scattering cross section of the probe \cite{Balistreri_01}, confirming that near-field microscopy measurements only perturb tightly confined fields.\par
	
	\begin{figure}[htb!]
		\centering{\includegraphics[width=0.75\textwidth]{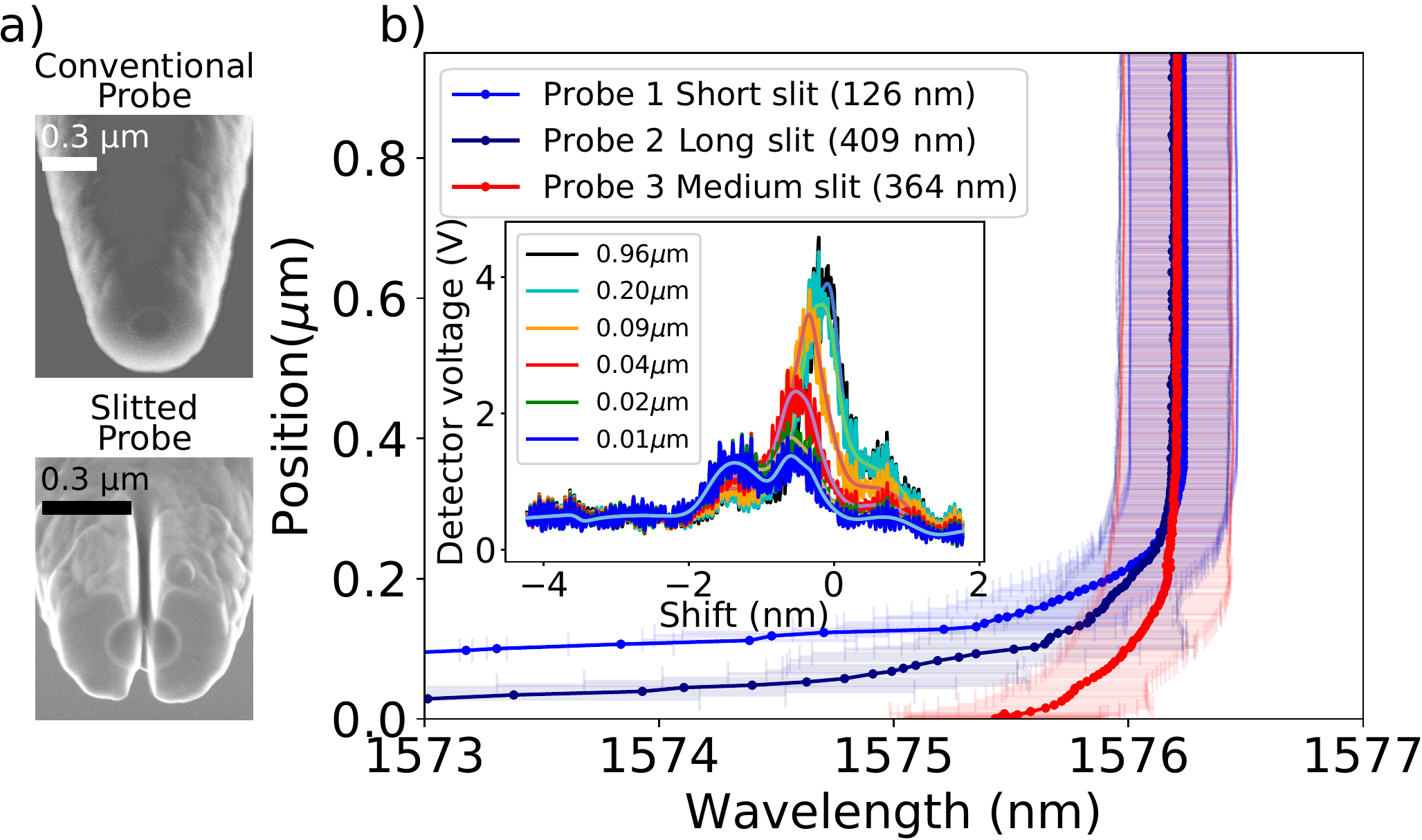}}
		\caption{Reduction of cavity resonance shift using nanostructured probes. (a) SEM pictures of a conventional probe (top) and a slitted nanostructured probe (bottom). Panel (b) shows the center wavelength of Lorentzian fits on transmission spectral measurements of the cavity while being perturbed by probes with different slit length. The horizontal bars associated with each curve indicate the line-width (FWHM) of the Lorentzian fits. The spectra were acquired while changing the vertical ($z$-axis) distance of the probes from the center of the cavity. Due to its short slit length, Probe 1 shows a similar shift as a conventional probe. The inset shows the measured transmission spectra through the cavity in the presence of Probe 3 at different heights (the solid lines show the fits used).}
		\label{figure3differentshifts}
	\end{figure}
	
	In order to assess the probe-sample interaction in more detail and test how the length of the slits affects the response, we carried out transmission measurements through the photonic-crystal cavity in the presence of multiple probes having different geometrical parameters. As in Figs.~\ref{figure2shifts}(a) and (b) above, we accurately move the probe vertically (along the $z$-axis) approaching the center of the cavity. The resulting transmission spectra as a function of probe height are subsequently fitted with a Lorentzian line-shape to extract the wavelength shift, as well as the change in the line-width of the cavity resonance. Fig.~\ref{figure3differentshifts}(a) depicts SEM images of a conventional and a slitted probe, while Fig.~\ref{figure3differentshifts}(b) shows tranmsission measurements in the presence of three different probes, with slit lengths equal to: 126~nm (Probe 1), 409~nm (Probe 2), and 364~nm (Probe 3). Additional geometrical parameters for these probes are detailed in the Appendix.
	Fig.~\ref{figure3differentshifts}(b) shows the central wavelength and line-width values retrieved from the Lorentzian fit of the transmission spectra for Probe 1 to 3. We see that a probe with a short slit (Probe 1) induces a similar shift in the wavelength of the cavity resonance as a conventional probe does, which indicates that a short slit does not balance the electric and magnetic polarizabilities of the probe, as qualitatively predicted by our numerical calculations in Fig.~\ref{mappingFields}. When the length of the slit is significantly increased, as for Probe 2, we observe a strong reduction of the probe-induced perturbation, with a wavelength shift reduced down to 23\% of the original value at a probe height of 100~nm, i.e., a change from a resonance wavelength of 1573.29~nm (for Probe 1) to 1575.53~nm (for Probe 2). Finally, for Probe 3, which has a shorter slit than Probe 2, we observe a further drastic reduction of the probe-cavity perturbation with a shift reduced to about 7\% of the original shift at 100~nm height, i.e., from a wavelength of 1573.29~nm to a wavelength of 1575.99~nm. Also, this probe creates a perturbation of less than 0.5~nm at 30~nm height. This behavior indicates that longer slits do not necessarily mean better polarizability compensation, and therefore better perturbation suppression, which is again qualitatively consistent with our numerical calculations in Figs.~\ref{mappingFields}(c)-(e). 
	Indeed, the optimal balancing of the electric and magnetic dipolar responses depends on the specific frequency of the polarizability resonance induced by the slits. As mentioned above, the frequency of this resonance is also tuned by the change in the aluminum coating thickness between the probes, as well as the Al$_{3}$O$_{2}$ content of the probes. In our case, Probe 3, which shows the smallest perturbation on the measured fields, turned out to have the best combination of slit length, width, and Aluminum coating thickness among all the probes fabricated and measured. While it is challenging to precisely control the fabrication parameters of these tips, most of the nanostructured probes with a slitted coating still exhibited significant cloaking effects, similar to Probe 2, hence demonstrating the robustness of the proposed approach.\par
	%
	
	\begin{figure*}[htb!]
		\centering{\includegraphics[width=1.00\textwidth]{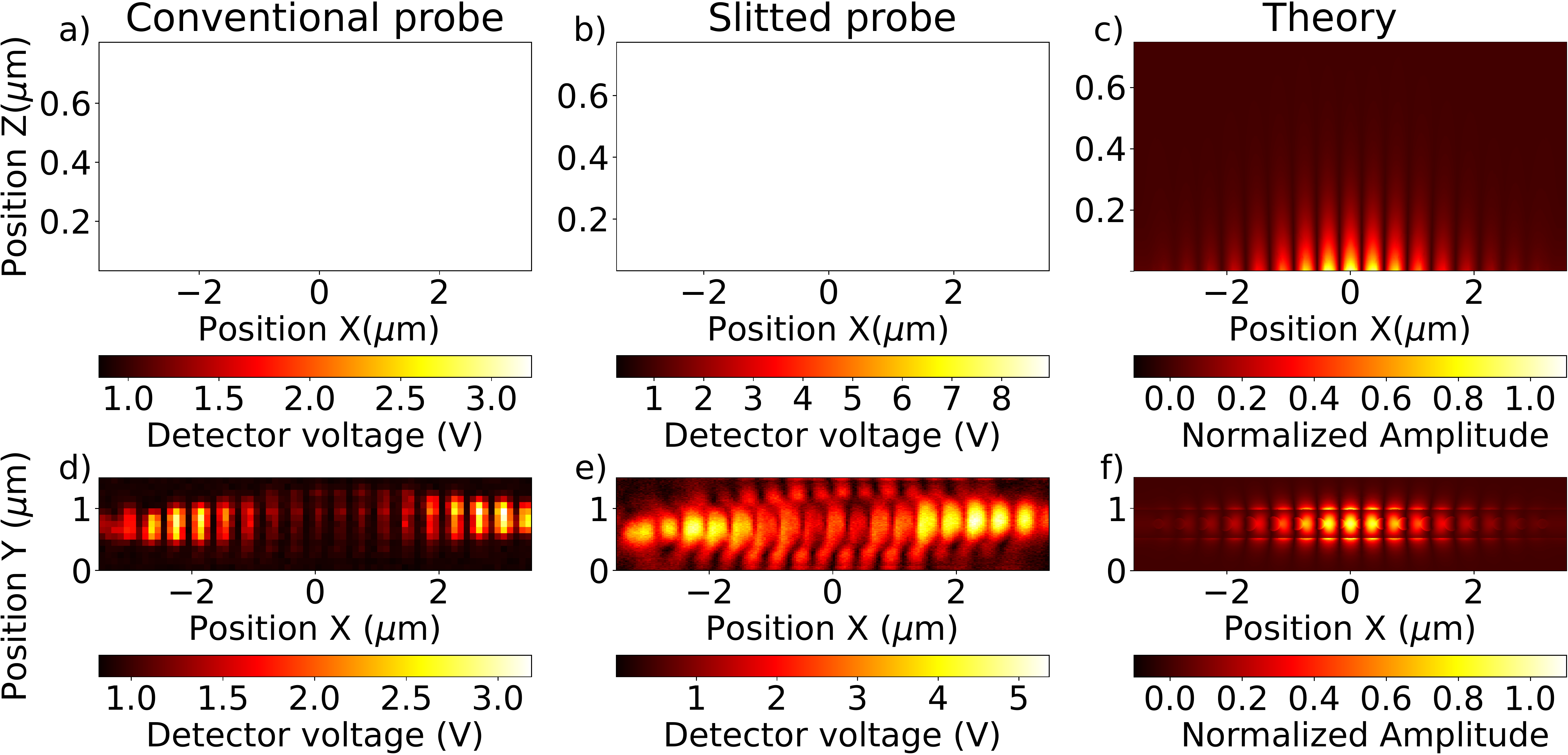}}
		\caption{Near-field amplitude maps measured through a near-field aperture probe, at a wavelength of 1576.2~nm, on the XZ plane (panels (a) and (b)) and the XY plane ((d) and (e)), using a conventional probe ((a) and (d)) and a nanostructured cloaked probe ((b) and (e)). The cloaked probe used here is Probe 3 in Fig.~\ref{figure3differentshifts}. These figures can be compared to the numerical calculations for the amplitude of E$_{y}$ in the resonant cavity, presented in panels ((c) and (f)).}\label{figure4fieldscavityZ}
	\end{figure*}

	The fact that a well balanced slitted probe reduces the resonance shift in transmission measurements through the nanophotonic cavity should also be reflected in much weaker perturbations on the measured near-field distribution. As a demonstration of the new capabilities of our ``cloaked'' probe, we present in Fig.~\ref{figure4fieldscavityZ} the electric near-field amplitude profiles of the nanophotonic cavity under study, mapped by a conventional unbalanced probe (Fig.~\ref{figure4fieldscavityZ}(a)~and~(d)) and a nanostructured balanced probe (Fig.~\ref{figure4fieldscavityZ}(b)~and~(e)). The measurements are carried out both in a vertical plane along the axis of mirror symmetry of the cavity (XZ plane), as well as in shear-force mode on the surface of the sample (XY plane). These measurements can be compared to the theoretical calculations presented in Fig.~\ref{figure4fieldscavityZ}(c)~and~(f) (note that these simulations, given their high amplitude at low heights close to the surface, show an apparently smaller field pattern extension due to saturation of the color scale). Similar to Fig.~\ref{mappingFields}(i), in Fig.~\ref{figure4fieldscavityZ}(a)~and~(d) we observe a region of lower field amplitude, where maximum intensity would be expected, caused by the perturbation of the conventional probe when it is in close proximity to the center of the cavity. This effect is clearly visible in both the XZ and the XY plane. The interaction of the probe with the cavity creates a strong contrast between the high amplitude fields at the edge of the cavity and the fields at the center of the cavity. For the conventional probe, the contrast between the maxima of the field antinodes away from the center and the maxima at the center of the cavity is 57\% at 100~nm height. Rather remarkably, our results in Figs.~\ref{figure4fieldscavityZ}(b)~and~(e) show that, when a nanostructured slitted probe with balanced electric and magnetic response is used (Probe 3 in Fig.~\ref{figure3differentshifts}), the amplitude contrast is strongly suppressed: only 16\% at 100~nm height, which is less than 30\% of the amplitude decrease observed with a conventional probe. These results clearly demonstrate the potential of the designed cloaked probes to achieve low levels of perturbation in near-field subdiffractive measurements.
	
	\section{Discussion}
	
	Inspired by the concepts of scattering-cancellation invisibility cloaks, cloaked sensors, and Huygens-Kerker scatterers, we have experimentally demonstrated a novel method to suppress the perturbations introduced by NSOM probes for near-field optical imaging at infrared frequencies, effectively making the probes invisible to the field distribution under study, without preventing their ability to measure. In order to characterize our nanostructured ``cloaked'' probes, we have implemented a heterodyne detection scheme, in which we simultaneously measure 3D raster-scanned field maps of the near-field amplitude distribution and the transmission through a nanophotonic cavity under study. In addition, in this measurement scheme, we have implemented a fast spectroscopic scan that allowed us to characterize the full electromagnetic response of the probes, in amplitude and phase, as a function of wavelength. 
	
	In conclusion, our experimental work demonstrates, for the first time, that a suitably nanostructured cloaked probe enables minimally-perturbative measurements of tightly confined resonant fields. In particular, we have shown that, by balancing the electric and magnetic dipole response of the near-field probes, we can reduce the perturbation by at least 70\%, quantified in terms of the resonance shift at probe-sample distances of $\sim$100~nm or less. This reduced perturbation allows mapping a field profile that closely resembles the eigenmodes of the cavity/waveguide under study, instead of the modes of the perturbed sample-probe system. Our results may pave the way for applying near-field microscopy to fragile perturbation-sensitive systems as in cavity quantum electrodynamics, as well as for research on single-emitter near-field dynamics. More generally, our work shows a path toward minimally-perturbative subdiffractive near-field imaging, and may lead to innovative advances in our exploration of the nano-world with optics.

	\section*{Acknowledgments}
	F.B.A., F.A., A.O and L.K. acknowledge funding from the European Research Council (ERC) Advanced Investigator Grant no. 340438-CONSTANS. This work is part of the research program of The Netherlands Organization for Scientific Research (NWO). F.M. and A.C. acknowledge support from the Air Force Office of Scientific Research with Grant No. FA9550-19-1-0043. The authors would like to thank Hinco Schoenmaker for technical support and fabrication of the probes and Dolfine Kosters for the initial measurements and setting up for this project.
	
	\section*{Disclosures}
	
	The authors declare no conflicts of interest.
	
	\section*{Data availability}
	
	Data underlying the results presented in this paper are not publicly available at this time but may be obtained from the authors upon reasonable request.
	
	\section*{Appendix: Methods}
	%

	\textit{Transmission measurements:} To find the spectral dependence of the resonance shift, we measure the transmission through the sample interferometrically with a heterodyne detection scheme (see Ref. \cite{Feber_2013}). In this scheme we combine the signal transmitted through the sample with a reference signal that has been phase-modulated, with an acousto-optical modulator, at a frequency of 40~kHz. Therefore, the combined light beams present a beating signal (at 40~kHz) that is measured with a lock-in technique, which yields the amplitude as well as the phase of the fields.\par
	\textit{Probe-sample distance control:} The probe-sample distance is kept constant through a shear-force feedback mechanism, as explained in chapter 7 of Ref.~\cite{Novotny2006}, which maintains the probe at a distance of around 10~nm with fluctuations of $\pm$2~nm. When the distance between the probe and the sample is bigger than $\sim$20~nm, it is necessary to use a different type of feedback mechanism. Specifically, we use the signal from a commercial interferometer (PicoScale GmbH) to control the probe positioning stage.\par
	\textit{Probes:} Probe 1 has a slit width of 30~nm, while the silica core is not cut through; it has a slit length equal to 126~nm, a silica core of 228~nm and an aluminum coating thickness of 179~nm. Probe 2 has a slit width of 50~nm and the silica core is cut through; it has a slit length of 409~nm, a silica core of 236~nm and an aluminum coating thickness of 160~nm. Probe 3 has a slit width of 36~nm, a slit length equal to 364~nm, a silica core of 163~nm and an aluminum coating thickness of 155~nm. The conventional probes used for comparison have a silica core of $\sim$240~nm and a coating thickness of $\sim$190~nm.\par 
	\textit{Complex Lorentzian line-shape:} Since in our setup we measure complex amplitudes and not intensities we use a complex Lorentzian-like line-shape for the fitting routine, given by the expression,
	\begin{equation}
		f(x)=b+\frac{A G e^{i \phi}}{1/x-1/x_{0}+i G},
	\end{equation}
	where $b$ is determined by the background level, $A$ is the amplitude of the peak, $G$ is related to the FWHM of the peak, and $x_{0}$ is the center of the peak.\par
	\textit{Simulations:} In order to obtain the polarizability tensor of the probe, we use a polarizability tensor retrieval method in which the scattered fields are projected onto vector spherical harmonics with the aid of an exact discrete spherical harmonic Fourier transform on the unit sphere. This method is explained in detail in Refs. \cite{Bernal_2013,Bernal_2014}. Full-wave eigenmode simulations of the nanophotonic cavity in Fig. \ref{figure4fieldscavityZ} have been performed in COMSOL Multiphysics. 

	\bibliography{biblioslittip}

\begin{thebibliography}{10}
\newcommand{\enquote}[1]{``#1''}

\bibitem{jackson1999}
J.~Jackson, \emph{Classical electrodynamics} (Wiley, New York, 1999).

\bibitem{bohren2008absorption}
C.~F. Bohren and D.~R. Huffman, \emph{Absorption and scattering of light by
  small particles} (John Wiley \& Sons, 2008).

\bibitem{Synge_1928}
E.~H. Synge, \enquote{A suggested method for extending microscopic resolution
  into the ultra-microscopic region,} {\protect\JournalTitle{The London,
  Edinburgh, and Dublin Philosophical Magazine and Journal of Science}}
  \textbf{6}, 356--362 (1928).

\bibitem{deLange2001}
F.~de~Lange, A.~Cambi, R.~Huijbens, B.~de~Bakker, W.~Rensen, M.~Garcia-Parajo,
  N.~van Hulst, and C.~G. Figdor, \enquote{Cell biology beyond the diffraction
  limit: near-field scanning optical microscopy,}
  {\protect\JournalTitle{Journal of Cell Science}} \textbf{114}, 4153--4160
  (2001).

\bibitem{Dorfmuller_2011}
J.~Dorfmüller, D.~Dregely, M.~Esslinger, W.~Khunsin, R.~Vogelgesang, K.~Kern,
  and H.~Giessen, \enquote{Near-field dynamics of optical yagi-uda
  nanoantennas,} {\protect\JournalTitle{Nano Letters}} \textbf{11}, 2819--2824
  (2011). PMID: 21619018.

\bibitem{Schnell2009}
M.~Schnell, A.~García-Etxarri, A.~J. Huber, K.~Crozier, J.~Aizpurua, and
  R.~Hillenbrand, \enquote{Controlling the near-field oscillations of loaded
  plasmonic nanoantennas,} {\protect\JournalTitle{Nature Photonics}}
  \textbf{3}, 287 (2009).

\bibitem{Balistreri_01}
M.~L.~M. Balistreri, J.~P. Korterik, L.~Kuipers, and N.~F. van Hulst,
  \enquote{Phase mapping of optical fields in integrated optical waveguide
  structures,} {\protect\JournalTitle{J. Lightwave Technol.}} \textbf{19}, 1169
  (2001).

\bibitem{Koenderink_2005}
A.~F. Koenderink, M.~Kafesaki, B.~C. Buchler, and V.~Sandoghdar,
  \enquote{Controlling the resonance of a photonic crystal microcavity by a
  near-field probe,} {\protect\JournalTitle{Phys. Rev. Lett.}} \textbf{95},
  153904 (2005).

\bibitem{Cognee_2019}
K.~G. Cogn\'{e}e, W.~Yan, F.~L. China, D.~Balestri, F.~Intonti, M.~Gurioli,
  A.~F. Koenderink, and P.~Lalanne, \enquote{Mapping complex mode volumes with
  cavity perturbation theory,} {\protect\JournalTitle{Optica}} \textbf{6},
  269--273 (2019).

\bibitem{Lalouat_2007}
L.~Lalouat, B.~Cluzel, P.~Velha, E.~Picard, D.~Peyrade, J.~P. Hugonin,
  P.~Lalanne, E.~Hadji, and F.~de~Fornel, \enquote{Near-field interactions
  between a subwavelength tip and a small-volume photonic-crystal nanocavity,}
  {\protect\JournalTitle{Phys. Rev. B}} \textbf{76}, 041102 (2007).

\bibitem{Intoti_2008}
F.~Intonti, S.~Vignolini, F.~Riboli, A.~Vinattieri, D.~S. Wiersma, M.~Colocci,
  L.~Balet, C.~Monat, C.~Zinoni, L.~H. Li, R.~Houdr\'e, M.~Francardi,
  A.~Gerardino, A.~Fiore, and M.~Gurioli, \enquote{Spectral tuning and
  near-field imaging of photonic crystal microcavities,}
  {\protect\JournalTitle{Phys. Rev. B}} \textbf{78}, 041401 (2008).

\bibitem{Vignolini_2010}
S.~Vignolini, F.~Intonti, F.~Riboli, L.~Balet, L.~H. Li, M.~Francardi,
  A.~Gerardino, A.~Fiore, D.~S. Wiersma, and M.~Gurioli, \enquote{Magnetic
  imaging in photonic crystal microcavities,} {\protect\JournalTitle{Phys. Rev.
  Lett.}} \textbf{105}, 123902 (2010).

\bibitem{Burresi_2010}
M.~Burresi, T.~Kampfrath, D.~van Oosten, J.~C. Prangsma, B.~S. Song, S.~Noda,
  and L.~Kuipers, \enquote{Magnetic light-matter interactions in a photonic
  crystal nanocavity,} {\protect\JournalTitle{Phys. Rev. Lett.}} \textbf{105},
  123901 (2010).

\bibitem{Fleury_2015}
R.~Fleury, F.~Monticone, and A.~Al\`u, \enquote{Invisibility and cloaking:
  Origins, present, and future perspectives,} {\protect\JournalTitle{Phys. Rev.
  Applied}} \textbf{4}, 037001 (2015).

\bibitem{Kerker_1975}
M.~Kerker, \enquote{Invisible bodies,} {\protect\JournalTitle{J. Opt. Soc.
  Am.}} \textbf{65}, 376--379 (1975).

\bibitem{Alu_2008}
A.~Al\`u and N.~Engheta, \enquote{Plasmonic and metamaterial cloaking: physical
  mechanisms and potentials,} {\protect\JournalTitle{Journal of Optics A: Pure
  and Applied Optics}} \textbf{10}, 093002 (2008).

\bibitem{Estakhri_2014}
N.~M. Estakhri and A.~Al\`u, \enquote{Ultra-thin unidirectional carpet cloak
  and wavefront reconstruction with graded metasurfaces,}
  {\protect\JournalTitle{IEEE Antennas and Wireless Propagation Letters}}
  \textbf{13}, 1775--1778 (2014).

\bibitem{Ni_2015}
X.~Ni, Z.~J. Wong, M.~Mrejen, Y.~Wang, and X.~Zhang, \enquote{An ultrathin
  invisibility skin cloak for visible light,} {\protect\JournalTitle{Science}}
  \textbf{349}, 1310--1314 (2015).

\bibitem{Schurig_06}
D.~Schurig, J.~J. Mock, B.~J. Justice, S.~A. Cummer, J.~B. Pendry, A.~F. Starr,
  and D.~R. Smith, \enquote{Metamaterial electromagnetic cloak at microwave
  frequencies,} {\protect\JournalTitle{Science}} \textbf{314}, 977--980 (2006).

\bibitem{Alu_2005}
A.~Al\`u and N.~Engheta, \enquote{Achieving transparency with plasmonic and
  metamaterial coatings,} {\protect\JournalTitle{Phys. Rev. E}} \textbf{72},
  016623 (2005).

\bibitem{Valentine_2009}
J.~Valentine, J.~Li, T.~Zentgraf, G.~Bartal, and X.~Zhang, \enquote{An optical
  cloak made of dielectrics,} {\protect\JournalTitle{Nature Materials}}
  \textbf{8}, 568 (2009).

\bibitem{Soric_14}
J.~C. {Soric}, R.~{Fleury}, A.~{Monti}, A.~{Toscano}, F.~{Bilotti}, and
  A.~{Al\`u}, \enquote{Controlling scattering and absorption with metamaterial
  covers,} {\protect\JournalTitle{IEEE Transactions on Antennas and
  Propagation}} \textbf{62}, 4220--4229 (2014).

\bibitem{chen2019active}
A.~Chen and F.~Monticone, \enquote{Active scattering-cancellation cloaking:
  broadband invisibility and stability constraints,}
  {\protect\JournalTitle{IEEE Transactions on Antennas and Propagation}}
  \textbf{68}, 1655--1664 (2019).

\bibitem{Alu_2009}
A.~Al\`u and N.~Engheta, \enquote{Cloaking a sensor,}
  {\protect\JournalTitle{Phys. Rev. Lett.}} \textbf{102}, 233901 (2009).

\bibitem{Fleury_14}
R.~Fleury, J.~Soric, and A.~Al\`u, \enquote{Physical bounds on absorption and
  scattering for cloaked sensors,} {\protect\JournalTitle{Phys. Rev. B}}
  \textbf{89}, 045122 (2014).

\bibitem{Alu_2010}
A.~Al\`u and N.~Engheta, \enquote{Cloaked near-field scanning optical
  microscope tip for noninvasive near-field imaging,}
  {\protect\JournalTitle{Phys. Rev. Lett.}} \textbf{105}, 263906 (2010).

\bibitem{Chen_2017}
A.~Chen, A.~Al\`u, and F.~Monticone, \enquote{Invisible near-field probes at
  infrared frequencies based on impedance engineering at the nanoscale,} in
  \emph{2017 IEEE International Symposium on Antennas and Propagation USNC/URSI
  National Radio Science Meeting,}  (2017), pp. 1055--1056.

\bibitem{Monticone_2014}
F.~Monticone and A.~Al\`u, \enquote{Embedded photonic eigenvalues in 3d
  nanostructures,} {\protect\JournalTitle{Phys. Rev. Lett.}} \textbf{112},
  213903 (2014).

\bibitem{Alukerker_2010}
A.~Al\`u and N.~Engheta, \enquote{How does zero forward-scattering in
  magnetodielectric nanoparticles comply with the optical theorem?}
  {\protect\JournalTitle{Journal of Nanophotonics}} \textbf{4}, 1 -- 18 -- 18
  (2010).

\bibitem{picardi2018janus}
M.~F. Picardi, A.~V. Zayats, and F.~J. Rodr{\'\i}guez-Fortu{\~n}o,
  \enquote{Janus and huygens dipoles: near-field directionality beyond
  spin-momentum locking,} {\protect\JournalTitle{Physical review letters}}
  \textbf{120}, 117402 (2018).

\bibitem{picardi2019experimental}
M.~F. Picardi, M.~Neugebauer, J.~S. Eismann, G.~Leuchs, P.~Banzer, F.~J.
  Rodr{\'\i}guez-Fortu{\~n}o, and A.~V. Zayats, \enquote{Experimental
  demonstration of linear and spinning janus dipoles for polarisation-and
  wavelength-selective near-field coupling,} {\protect\JournalTitle{Light:
  Science \& Applications}} \textbf{8}, 1--7 (2019).

\bibitem{Conteduca_2017}
D.~Conteduca, C.~Reardon, M.~G. Scullion, F.~Dell’Olio, M.~N. Armenise, T.~F.
  Krauss, and C.~Ciminelli, \enquote{Ultra-high q/v hybrid cavity for strong
  light-matter interaction,} {\protect\JournalTitle{APL Photonics}} \textbf{2},
  086101 (2017).

\bibitem{Novotny2006}
L.~Novotny, \emph{Principles of Nano-Optics} (Cambridge University Press,
  2006).

\bibitem{fleischer2012near}
M.~Fleischer, \enquote{Near-field scanning optical microscopy nanoprobes,}
  {\protect\JournalTitle{Nanotechnology Reviews}} \textbf{1}, 313--338 (2012).

\bibitem{Sersic_2009}
I.~Sersic, M.~Frimmer, E.~Verhagen, and A.~F. Koenderink, \enquote{Electric and
  magnetic dipole coupling in near-infrared split-ring metamaterial arrays,}
  {\protect\JournalTitle{Phys. Rev. Lett.}} \textbf{103}, 213902 (2009).

\bibitem{Bernal_2013}
F.~Bernal~Arango and A.~F. Koenderink, \enquote{Polarizability tensor retrieval
  for magnetic and plasmonic antenna design,} {\protect\JournalTitle{New
  Journal of Physics}} \textbf{15}, 073023 (2013).

\bibitem{Olmon_2010}
R.~L. Olmon, M.~Rang, P.~M. Krenz, B.~A. Lail, L.~V. Saraf, G.~D. Boreman, and
  M.~B. Raschke, \enquote{Determination of electric-field, magnetic-field, and
  electric-current distributions of infrared optical antennas: A near-field
  optical vector network analyzer,} {\protect\JournalTitle{Physical Review
  Letters}} \textbf{105}, 167403 (2010).

\bibitem{Devaux_2000}
E.~Devaux, A.~Dereux, E.~Bourillot, J.-C. Weeber, Y.~Lacroute, J.-P. Goudonnet,
  and C.~Girard, \enquote{Local detection of the optical magnetic field in the
  near zone of dielectric samples,} {\protect\JournalTitle{Phys. Rev. B}}
  \textbf{62}, 10504--10514 (2000).

\bibitem{Kihm_2011}
H.~W. Kihm, S.~M. Koo, Q.~H. Kim, K.~Bao, J.~E. Kihm, W.~S. Bak, S.~H. Eah,
  C.~Lienau, H.~Kim, P.~Nordlander, N.~J. Halas, N.~K. Park, and D.-S. Kim,
  \enquote{Bethe-hole polarization analyser for the magnetic vector of light,}
  {\protect\JournalTitle{Nature Communications}} \textbf{2}, 451 (2011).

\bibitem{Ribaric_95}
M.~Ribarič and L.~Šušteršič, \enquote{Expansions in terms of moments of
  time-dependent, moving charges and currents,} {\protect\JournalTitle{SIAM
  Journal on Applied Mathematics}} \textbf{55}, 593--624 (1995).

\bibitem{smajic2011numerical}
J.~Smajic and C.~Hafner, \enquote{Numerical analysis of a snom tip based on a
  partially cladded optical fiber,} {\protect\JournalTitle{Optics express}}
  \textbf{19}, 23140--23152 (2011).

\bibitem{qian2015plasmonic}
Q.~Qian, H.~Yu, P.~Gou, J.~Xu, and Z.~An, \enquote{Plasmonic focusing of
  infrared snom tip patterned with asymmetric structures,}
  {\protect\JournalTitle{Optics express}} \textbf{23}, 12923--12934 (2015).

\bibitem{wei2018directional}
L.~Wei, M.~F. Picardi, J.~J. Kingsley-Smith, A.~V. Zayats, and F.~J.
  Rodr{\'\i}guez-Fortu{\~n}o, \enquote{Directional scattering from particles
  under evanescent wave illumination: the role of reactive power,}
  {\protect\JournalTitle{Optics letters}} \textbf{43}, 3393--3396 (2018).

\bibitem{Bernal_2014}
F.~Bernal~Arango, T.~Coenen, and A.~F. Koenderink, \enquote{Underpinning
  hybridization intuition for complex nanoantennas by magnetoelectric
  quadrupolar polarizability retrieval,} {\protect\JournalTitle{ACS Photonics}}
  \textbf{1}, 444--453 (2014).

\bibitem{Feber_2013}
B.~le~Feber, N.~Rotenberg, D.~M. Beggs, and L.~Kuipers, \enquote{Simultaneous
  measurement of nanoscale electric and magnetic optical fields,}
  {\protect\JournalTitle{Nature Photonics}} \textbf{8}, 43 (2013).

\end{thebibliography}
\end{document}